\documentstyle[epsf,psfig,11pt]{article}
\newcommand{\be}{\begin{eqnarray}}

\newcommand{\ee}{\end{eqnarray}}
\newcommand{\munu}{\mu\nu}
\newcommand{\psibar}{\bar \psi}
\newcommand{\nhat}{\hat{n}}

\title{
\bf  Classical methods in DIS and nuclear scattering at small x}
\author{ Raju Venugopalan\\
        {\small\it Physics Department,
        Brookhaven National Laboratory,
        Upton, NY 11973, USA } \\
         }
\date{}

\begin{document}

\maketitle

\begin{center}
{\bf Abstract}\\
\end{center}
In hadrons and nuclei at very small $x$, parton distributions saturate
at a scale $Q_s(x)$.  Since the occupation number is large, and
$Q_s(x)>>\Lambda_{QCD}$, classical weak coupling methods may be used
to study this novel regime of non--linear classical fields in QCD. In
these lectures, we apply these methods to compute structure functions
in deeply inelastic scattering (DIS) and the energy density of gluons
produced in high energy nuclear collisions.

\vfill \eject

\section{Introduction}

One of more interesting problems in perturbative QCD is the behaviour
of parton distributions at small values of Bjorken $x$. In deeply inelastic
scattering (DIS) for instance, for a fixed $Q^2>>\Lambda_{QCD}^2$, the
operator product expansion (OPE) eventually breaks down at sufficiently small
x~\cite{AMueller1}. Therefore at asymptotic energies, the conventional
approaches towards computing observables based on the linear
DGLAP~\cite{DGLAP} equations are no longer applicable. 
Even at current collider energies such as 
those of HERA, where the conventional wisdom is that the DGLAP equations
successfully describe the data, there is reason to believe that
effects due to large
parton densities are not small. We may be at the threshold of a
region where non--linear corrections to the evolution equations
are large~\cite{LanLev,FrankStrik1}.

In recent years, a non--OPE based effective field theory approach to
small $x$ physics has been developed by Lipatov and
collaborators~\cite{Lipatov}. Their initial efforts resulted in an
equation known popularly as the BFKL equation~\cite{BFKL}, which sums
the leading logarithms of $\alpha_S\log(1/x)$ in QCD. In marked
contrast to the leading twist Altarelli--Parisi equations for
instance, it sums all twist operators that contain the leading
logarithms in x. The solutions to the BFKL equation predict a rapidly
rising gluon density. Such a rapid rise in the gluon density is seen at 
HERA~\cite{H1ZEUS} but it can also arguably be accounted for
by the next to leading order (NLO) 
DGLAP equations with appropriate choices of the initial
parton densities~\cite{GRVCTEQ}.

Moreover, the next to leading logarithmic corrections to the BFKL equation
computed in the above mentioned effective field theory (EFT) approach are
{\it very} large~\cite{FadLip}.  Recently, as Gavin
Salam has discussed in his lectures at this school~\cite{Gavin}, 
there has been considerable
progress in understanding the source of these large corrections. The
collinear enhancements from higher orders to the NLO BFKL kernel can be
resummed, and this results in more stable estimates for the gluon
anomalous dimensions, and for the hard pomeron. However,
there are effects, not included in such analyses, due to multiple
pomeron exchange (non--linear QCD effects) that may become important
at rapidity scales of interest. For instance, running coupling effects
in the NLO BFKL equation become important at $y\sim 1/\alpha_S^{5/3}$.
However, double (hard) pomeron effects will become important for
$y\sim 1/\alpha_S \log (1/\alpha_S)$, a scale that is parametrically
larger.  How to systematically include such effects, which are
enhanced by large parton densities, is an open question, and novel
approaches need to be explored.

An alternative EFT approach to QCD at small $x$ was
put forward in a series of papers~\cite{MV,AJMV,JKMW,JKLW,JKW}.  
Our approach, is a
Wilson renormalization group approach (RG) where the fields are those
of the fundamental theory but the form of the action, at small $x$, is
obtained by integrating out modes at higher values of $x$. This 
results in a set of non--linear renormalization
group equations~\cite{JKW}. If the parton densities are not too high,
the RG equations can be linearized, and have been
shown to agree identically with the BFKL and small $x$ DGLAP
equations~\cite{JKLW}. There is much effort underway to explore and
make quantitative predictions for the non--linear regime
beyond~\cite{JKW2ILM}.

In these lectures, we will apply the above EFT
to discuss two problems:
\begin{itemize}
\item 
Deeply inelastic scattering at small values of Bjorken $x$~\cite{MV99},
\item 
high energy hadronic collisions~\cite{AlexRaj1,AlexRaj2,AlexRaj3}.
\end{itemize}
Both are problems which appear extremely difficult to address in an OPE
based analysis. They simplify in the regime 
where $x$ is small and momentum transfers 
are large because, a) the gluon field at small
$x$ can be treated classically, and b) weak coupling methods apply.
Why is this so?  The reasons for these to apply have been discussed at
length by Larry McLerran in his lectures~\cite{Larrylectures} so we
will be brief here.

At small $x$, one can define a scale $\mu^2$
which measures the density of gluons per unit transverse area. One has 
\be
\mu^2 = {1 \over \sigma} {{dN} \over {dy}} \, 
\label{eq:1}
\ee
where $\sigma$ is the hadronic or nuclear cross section of interest.
Here $y =y_0 - ln(1/x)$, and $y_0$ is an arbitrarily chosen constant.  
When this scale satisfies the condition $\mu\gg
\alpha_S\mu \gg\Lambda_{QCD}$, the occupation number of gluons in the
hadron is large--thereby justifying the use of classical methods. Also, 
the intrinsic momentum $p_t\sim \alpha_S\mu$ is large. Thus the gluon
dynamics, while nonperturbative, is both semiclassical and weakly
coupled.

Let us now discuss the classical field approach to small $x$ DIS. 
The gluon field, being bosonic, has to be treated
non-perturbatively.  This is analogous to the strong field limit used
in Coulomb problems.  Fermions, on the other hand, do not develop a
large expectation value and may be treated perturbatively.  In DIS, to
lowest order in $\alpha_S$, the gluon distribution function is
determined by knowing the fermionic propagator in the classical gluon
background field.  In general, this propagator must be determined to
all orders in the classical gluon field as the field is strong.  This
can be done due to the simple structure of the background field.

We will derive analytic expressions for the current--current
correlator in deeply inelastic scattering by summing a particular
class of all twist operators. These, we argue, give the dominant
contribution at small $x$. At high $Q^2$, they reduce to the well
known expressions for the leading twist structure
functions~\cite{Jaffe}.  For light quarks at high $Q^2$, it can be
shown explicitly that the classical field analysis reproduces the
DGLAP evolution equations for the quark distributions at small
$x$~\cite{MV99}.  The power of the technique we use to analyze the
problem of DIS at small $x$ is that, unlike the OPE, it does not rely
on a twist expansion.

A similar point can be made about the classical approach to high
energy hadronic scattering.  At very high energies, the dominant
contribution to particle production is from the interaction of the
classical ``Weizs\"{a}cker--Williams'' (WW) gluon fields of the two
hadrons or nuclei.  To lowest order, the picture is that of QCD
Bremsstrahlung~\cite{GunionBertsch}. Soft gluons can be emitted from
the valence quark/hard gluon lines, or from the 2$\rightarrow$1
diagram of two virtual gluons fusing to produce a hard gluon. At small
$x$, in the Fock--Schwinger gauge ($x^+ A^- + x^- A^+ =0$) the latter
WW contribution is the dominant one. The Weizs\"{a}cker--Williams
contribution agrees with the QCD Bremsstrahlung result at small
$x$~\cite{KLW,gyulassy,DirkYuri,SerBerDir,Guo}.

It is essential to consider the full non--perturbative approach for
the following reasons. Firstly, the classical gluon radiation computed
perturbatively is infrared singular, and has to be cut-off at some
scale $k_t\sim \alpha_S\mu$.  This problem also arises in mini--jet
calculations where at high energies results are shown to be rather
sensitive to the cut--off~\cite{minijets}.  Secondly, the
non--perturbative approach is crucial to a study of the space--time
evolution of the system. In particular, the possible thermalization of
the system, as well as the relevant time scales for thermalization,
are strongly influenced by the non--linearities that arise in the
non--perturbative approach~\cite{Muell}.
 
In these lectures, we will discuss results from real time simulations
of the full, non--perturbative, evolution of classical non--Abelian WW
fields. The fields are generated by sources of color charge $\rho^\pm$
(representing the valence partons in each of the hadrons or nuclei)
moving along the two light cones.  For each $\rho$ configuration, one
solves Hamilton's equation numerically to obtain the real time
behavior of the gauge fields in the forward light cone. The
Hamiltonian is the Kogut--Susskind Hamiltonian in 2+1--dimensions
coupled to an adjoint scalar field.  The initial conditions for the
evolution are provided by the non--Abelian Weizs\"acker--Williams
fields for the nuclei before the collisions.
 
To compute observables, one has to average over all the $\rho$ configurations 
in each of the two nuclei. In general, these are averaged with 
a statistical weight $\exp\left[-F[\rho]\right]$,
where $F[\rho]$ is a functional over the color charge density $\rho$
of the higher $x$ modes. The functional $F[\rho]$ obeys the non--linear
renormalization group equation that was mentioned in the preceding discussion. 
If one considers collisions of large nuclei, the weight simplifies to 
a Gaussian one, and one can replace
\be
F[\rho^\pm]\longrightarrow \int\, d^2 x_t\int_y^{y_{frag}^\pm}\,dy^\prime\,
{1\over \mu^2(y^\prime)}\,{\rm Tr}\left((\rho^\pm)^2\right) \, ,
\ee
where $y_{frag}^\pm$ are the rapidities corresponding to the fragmentation 
regions of the two nuclei.

Our approach is limited because it is classical. 
However, if the effective action approach captures the
essential physics of the small $x$ modes of interest, then in the spirit of the
Wilson renormalization group, quantum information from the large $x$ modes
(above the rapidity of interest) is contained in the parameter $\mu^2$
discussed above, which grows rapidly as one goes to smaller $x$'s.
In principle, this information can be included in the 
classical lattice simulations.

The plan of these lectures is as follows. We will begin in section 2
by reviewing the effective action for the small $x$ modes in QCD.  We
also discuss the classical saddle point solutions of this effective
action. In section 3, we will discuss how one computes quark
production in the classical gluon background field. At small $x$, this
gives the dominant contribution to the structure functions measured in
deeply inelastic scattering. Structure functions are computed in
section 4. At high $Q^2$, our results reproduce the small $x$ DGLAP
results. For smaller values of $Q^2$, for Gaussian sources, one
obtains the Glauber result for the structure functions in agreement
with previous derivations in the nuclear rest frame.  Subsequent
sections concern gluon production in 
high energy scattering of (in particular) large
nuclei. The classical approach to the two--nucleus problem is
discussed in section 5. It is very hard to solve for the non--perturbative 
dynamics analytically. It has not yet been done. Instead, we derive a 
numerical algorithm which captures the essential 
physics of the two--nucleus problem. Results from
our numerical simulations are discussed in section 6. Section 7
summarizes the material contained in the lectures and outlines
directions of future research.

\section{Effective Field Theory for Small x Partons in QCD}
\vskip 0.1in

In the infinite momentum frame (IMF) $P^+\rightarrow \infty$, the effective
action for the soft modes of the gluon field with longitudinal momenta
$k^+<<P^+$ (or equivalently $x\equiv k^+/P^+ << 1$) can be written in
light cone gauge $A^+=0$ as
\be
S_{eff} &=& -\int d^4 x {1\over 4} G_{\munu}^{a}G^{\munu,a} +{i\over N_c}
\int d^2 x_t dx^- \rho^a(x_t,x^-)
{\rm Tr}\left(\tau^a W_{-\infty,\infty}[A^-](x^-,x_t)\right)\nonumber \\
&+& i\int d^2 x_t dx^- F[\rho^a(x_t,x^-)] \, .
\label{action}
\ee
Above, $G_{\munu}^a$ is the gluon field strength tensor, $\tau^a$ are
the $SU(N_c)$ matrices in the adjoint representation and $W$ is the
path ordered exponential in the $x^+$ direction in the adjoint
representation of $SU(N_c)$,
\be
W_{-\infty,\infty}[A^-](x^-,x_t) = P\exp\left[-ig\int dx^+
A_a^-(x^-,x_t)\tau^a\right] \, .
\ee
The action is a gauge invariant form~\cite{JKLW} of the
action that was proposed in Ref.~\cite{MV}. One can write an alternative 
gauge invariant form of the action but the results are the same for the 
problem of interest.

The effective action considered here is valid in a limited range of
$P^+ << \Lambda^+$, where $\Lambda^+$ is an ultraviolet cutoff in the
plus component of the momentum.  The degrees of freedom at higher
values of $P^+$ have been integrated out. Their effect is to generate
the second and third terms in the action.  The first term is the usual
field strength piece of the QCD action and describes the dynamics of
the wee partons. The second term is the coupling of the wee partons to
the hard color charges at higher rapidities, with $x$ values
corresponding to values of $P^+ \ge \Lambda^+$.  When expanded to
first order in $A^-$, this term gives the familiar $J \cdot A$
coupling for Abelian classical fields. The last term in the effective
action is imaginary. It can be thought of as a statistical weight
resulting from integrating out the higher rapidity modes in the
original QCD action. Expectation values of gluonic operators $O(A)$
are then defined as
\be
<O(A)> = { \int [d\rho] \exp\left(-F[\rho]\right) \int [dA] O(A)
\exp\left(iS[\rho,A]\right) \over{\int [d\rho] \exp\left(-F[\rho]\right)
\int [dA] \exp\left(iS[\rho,A]\right)}} \, ,
\label{expvalue}
\ee
where $S[\rho,A]$ corresponds to the first two terms in
Eq.~\ref{action}.

In the IMF, only the $J^+$ component of the
current is large (the other components being suppressed by
$1/P^+$). The longer wavelength wee partons do not resolve the higher
rapidity parton sources to within $1/P^+$ and, for all practical
purposes, one may write 
\be
\rho^a (x_t,x^-)\longrightarrow \rho^a (x_t) \delta(x^-) \, .
\ee

In Ref.~\cite{MV} a Gaussian form for the action
\be
\int d^2 x_t {1\over 2\mu^2} \rho^a \rho^a \, ,
\label{Gauss}
\ee
was proposed, where $\mu^2$ was the average color charge squared per
unit
area of the sources at higher rapidities. 
For large nuclei $A>>1$ it was shown
that
\be
      \mu^2= {1 \over { \pi R^2}} {N_q \over {2N_c}} \sim A^{1/3}/6
\,\,\mbox{fm}^{-2}.
\ee 
This result was independently confirmed in a
model constructed in the nuclear rest frame~\cite{Kovchegov}.
If we include the contribution of gluons which have been integrated out by the
renormalization group technique, one finds that~\cite{gyulassy}
\be
        \mu^2 = {1 \over {\pi R^2}} \left( {N_q \over {2N_c} }+ {{N_cN_g} \over
{N_c^2-1}} \right)
\label{colordensity}
\ee
Here $N_q$ is the total number of quarks with x above the cutoff; 
$N_q = \sum_i \int_x^1 dx^\prime q_i(x^\prime)$
where the sum is over different flavors, spins, quarks and antiquarks. 
For gluons, we also have $N_g = \int_x^1 dx^\prime g(x^\prime)$. 
The value of $\pi R^2$ is well defined for a large nucleus.  For a
smaller hadron, we must take it to be $\sigma$, the total cross section
for hadronic interactions at an energy corresponding to the cutoff. 
This quantity will become better defined for a hadron in the
renormalization group analysis.   

The above equation for $\mu^2$ is subtle because, implicitly, on the right
hand side, there is a dependence on $\mu$ through the structure functions
themselves.  This is the scale at which they must be evaluated. 
Calculating $\mu$ therefore involves solving an implicit equation.
Note that because the gluon distribution function rises rapidly
at small x, the value of $\mu$ grows as x decreases. At some critical 
$x$, the indications are that the parton distributions saturate. Thus there 
may be a critical line in the $x$--$Q^2$ plane corresponding to parton 
saturation. This is an important 
point and we will return to it later.

The Gaussian
form of the functional $F[\rho]$ is reasonable when the color charges at
higher rapidity are uncorrelated and are random sources of color charge.
This is true for instance in a very large nucleus.  It is also true if
we study the Fock space distribution functions or deep inelastic
structure functions at a transverse momentum scale which is larger than
an intrinsic scale set by $\alpha_S \mu$.  In this equation
$\alpha_S$ is evaluated at the scale $\mu$.  At smaller transverse
momenta scales, one must do a complete renormalization group analysis to
determine $F[\rho]$. For heavy quarks, in DIS, the Gaussian analysis should 
be adequate.

In Ref.~\cite{JKMW}, it was shown that a Wilson renormalization group
procedure could be applied to derive a non-linear renormalization
group equation for $F[\rho]$. In the limit of weak fields, the
renormalization group equation can be linearized, and can be shown to
be none other than the BFKL equation discussed previously. The fact
that this limit can be obtained in a simple and elegant way suggests
the power of this approach, and the importance of further studying the
non--linear region of strong classical fields.  We will not discuss
the RG procedure here but will refer the reader to the relevant
papers, and to Larry McLerran's lectures~\cite{Larrylectures}.

The effective action in Eq.~\ref{action} has a remarkable saddle point
solution~\cite{MV,JKMW,Kovchegov}. It is equivalent to solving the Yang--Mills
equations
\be
D_\mu G^{\munu} = J^\nu \delta^{\nu +} \, ,
\ee
in the presence of the source $J^{+,a} = \rho^a(x_t,x^-)$.
Here we will allow the source to be smeared out in $x^-$ as this is
useful in the renormalization group analysis. It is also useful for intuitively
understanding the nature of the field.
One finds a solution where $A^{\pm}=0$ and
\be
A^i = {-1\over {ig}}\,V\partial^i V^\dagger  \, ,
\label{ap1}
\ee
($i = 1,2$)
is a pure gauge field in the two transverse dimensions which satisfies the equation
\be
D_i {d A^i\over dy} = g \rho (y,x_\perp) \, .
\label{ap2}
\ee 
Here $D_i$ is the covariant derivative $\partial_i + V\partial_i
V^\dagger$
and $y=y_0 + \log(x^-/x_0^-)$ is the space--time rapidity and $y_0$ is
the
space-time rapidity of the hard partons in the fragmentation region. At
small $x$, we will use the space--time and momentum space notions of
rapidity interchangeably~\cite{RajLar}.  The momentum space rapidity is
defined to be $y = y_0 - ln(1/x)$.
The solution of the above equation is
\be
A_\rho^i(x_t) = {1\over ig}\left(Pe^{ig\int_y^{y_0}
dy^\prime
{1\over {\nabla_{\perp}^2}}\rho(y^\prime,x_t)}\right)
\nabla^i\left(Pe^{ig\int_y^{y_0} dy^\prime {1\over
{\nabla_{\perp}^2}}\rho(y^\prime,x_t)}\right)^\dagger \, .
\label{puresoln}
\ee

The classical nuclear gluon distribution function is computed by  
averaging over the product of the classical fields
in Eq.~\ref{puresoln} at two space--time points with the weight
$F[\rho]$~\cite{MV}. For a Gaussian source, one obtains 
\be
{dN\over d^2 x_t} = {1\over 2\pi\alpha_S} {C_F\over 
x_t^2} \left( 1-\exp\left( -{\alpha_S\pi^2\over 2\sigma C_F} x_t^2 xG\left(x,{1
\over x_t^2}\right)\right) \right)\, ,
\label{glue}
\ee 
where $C_F$ is the Casimir in the fundamental representation and 
$\sigma$ is the nuclear cross-section\footnote
{ Above, we have re-written the expression for the gluon distribution
in Ref.~\cite{JKMW}, using the leading log gluon distribution to
replace $\mu^2$ and $\log(x_t\Lambda_{QCD})$ with the gluon distribution 
$xG(x,{1\over x_t^2})$ at the scale $1/x_t^2$.}.
For large $x_t$ (but smaller that $1/\Lambda_{QCD}$, the distribution
falls like a power law $1/x_t^2$--and has a $1/\alpha_S$ dependence! 
For very small $x_t$, the behavior
is the perturbative distribution $\log(x_t \Lambda_{QCD})$. The scale
which determines the cross--over from a logarithmic to a power law
distribution is, following Mueller's notation~\cite{Muell991}, the
saturation scale $Q_s$. Setting $x_t=1/Q_s$ and the argument of the 
exponential above to unity, one obtains the relation, 
\be
Q_s^2 = {\alpha_S \pi^2\over 2\sigma}{1\over C_F} xG(x,Q_s^2) \, ,
\label{gluesat}
\ee
which, for a particular $x$, can be solved self--consistently to
determine $Q_s$.

Because of the sharp cut-off in co--ordinate space, the momentum space
distribution is not well defined. A smooth Fourier transform has been
defined, on physical grounds, by Lam and Mahlon by requiring that the
charge in light cone gauge $\int d^2 x_t \rho(x_t)$
vanish~\cite{LamMahlon} for each $\rho$ configuration.

\section{Quark production in the classical gluon background field}
\vskip 0.1in

In this section, we will compute the correlator of 
electromagnetic currents in the classical gluon background field.
In deeply inelastic electroproduction, the hadron tensor can be expressed
in terms of the forward Compton scattering amplitude $T_{\munu}$ by
the relation~\cite{Pokorski}
\be
W^{\munu}(q^2,P\cdot q) &=& 2 Disc~ T^{\munu}(q^2,P\cdot q) \equiv
{1\over {2\pi}} {\rm Im} \int d^4 x \exp(iq\cdot x) \nonumber \\
&\times&  <P|T(J^\mu (x) J^\nu (0))|P> \, ,
\label{hadtensor}
\ee
where ``T'' denotes time ordered product, $J^\mu={\psibar}\gamma^\mu
\psi$ is the hadron electromagnetic current and ``Disc'' denotes the
discontinuity of $T_{\munu}$ along its branch cuts in the variable
$P\cdot q$.  Also, $q^2\rightarrow \infty$ is the momentum transfer
squared of the virtual photon\footnote{ Note that in our metric
convention, a space--like photon has $q^2 = Q^2 > 0$.} and $P$ is the
momentum of the target.  In the IMF, $P^+\rightarrow \infty$ is the
only large component of the momentum.  The fermion state above is
normalized as $<P \mid P^\prime> = (2\pi)^3 E/m \,\delta^{(3)}
(P-P^\prime)$ where $m$ is the mass of the target hadron. This
definition of $W^{\mu \nu}$ and normalization of the state is
traditional. In the end, all factors of $m$ cancel from the definition
of quantities of physical interest.  (The normalization we will use in
this paper for quark and lepton states will have E/m replaced by
$2P^+$.)

We now generalize our definition of 
$W^{\mu\nu}$ to a source which has a position dependence. We obtain  
\be
        W^{\mu \nu} (q^2, P\cdot q) = {1\over {2\pi}}\sigma {P^+ \over m} 
{\rm Im}  \int d^4x dX^- e^{i q\cdot x}
 < T\left(J^\mu(X^- +x/2) J^\nu (X^- -x/2)\right)>\, .\nonumber\\
\ee
To see this, first note that we can define $< O > = <P \mid O \mid P>
/ <P \mid P>$ where $O$ is any operator.  From the discussion above,
the expectation value $<P \mid P> = (2\pi)^3 E/m\, \delta^{(3)} (0) =
(2\pi)^3 E/m ~V$. Here we shall take the spatial volume $V$ to be
$\sigma$ times an integral over the longitudinal extent of the state.
The variable $X^-$ is a center of mass coordinate and $x^-$ is the
relative longitudinal position.  The above definition of $W^{\mu\nu}$
is Lorentz covariant.  The integration over $X^-$ is required since we
must include all of the contributions from quarks at all $X^-$ to the
distribution function.  In the external source language, the variable
$P^+$ can be taken to be the longitudinal momentum corresponding to
the fragmentation region.
 
The expectation value is straightforward to compute in the limit where
the gluon field is treated as a classical background field.  If we write
\be
  < T(J^\mu (x) J^\nu(y))> = <T\left(\overline \psi (x) \gamma^\mu \psi(x)
\overline \psi(y) \gamma ^\nu \psi(y)\right) >\, ,
\ee
then when the background field is classical, we obtain
\be
        <T(J^\mu(x) J^\nu(y))> = {\rm Tr}
(\gamma^\mu S_A(x)) {\rm Tr} (\gamma^\nu S_A(y)) +
{\rm Tr}(\gamma^\mu S_A(x,y) \gamma^\nu S_A(y,x) ) \, .
\label{tad}
\ee
In this expression, $S_A(x,y)$ is the Green's function for the fermion
field in the external field $A$
\be
        S_A(x,y) = -i< \psi(x) \overline \psi (y) >_A \, .
\label{propdef}
\ee

The first term on the right hand side of Eq.~\ref{tad} is a tadpole
contribution without an imaginary part.  It
therefore does not contribute to $W^{\mu\nu}$.  We find then that
\be
        W^{\mu \nu}(q^2,p\cdot q) &=& {1\over 2\pi}\sigma {P^+ \over m} 
{\rm Im} \int dX^- d^4x \,e^{iq\cdot x}\,
\langle{\rm Tr}\Big( \gamma^\mu S_A (X^- + x/2,X^- - x/2)\nonumber \\
&\times& \gamma^\nu 
S_A(X^- - x/2,X^- + x/2)\Big)\rangle \, .
\ee

The expression we derived above for $W^{\munu}$ is
entirely general and makes no reference to the operator product
expansion.  In particular, it is relevant at the
small $x$ values and moderate $q^2$ where the operator product expansion
is not reliable~\cite{AMueller1}.  At sufficiently high $q^2$ though
(and for massless quarks) it should agree with the usual leading twist
computation of the structure functions.

We can derive an expression for the
sea quark  Fock distribution in terms of the propagator in light
cone quantization~\cite{KogutSoper}.  
One obtains 
\be
{dN\over {d^3 k}} = {2 i\over {(2\pi)^3}}\int d^3 x\, d^3 y\,
e^{-ik\cdot (x-y)}\,{\rm Tr}\left[\gamma^+ S_A(x,y)\right] \,\,
\label{distprop}
\ee
where the fermion propagator $S_A(x,y)$ is defined as in Eq.~\ref{propdef}. 
In a nice pedagogical paper, (see Ref.~\cite{Jaffe} and references
within),
Jaffe has shown that the Fock space distribution function can be
simply related to the {\it leading twist} 
structure function $F_2$ by the relation
\be
F_2(x,Q^2) = \int_0^{Q^2} dk_t^2 x {dN\over{dk_t^2 dx}}  \, .
\label{ltf2}
\ee
Actually, Jaffe's expression is defined as the sum of the quark and
anti--quark distributions. At small $x$, these are identical and the
resulting factor of 2 is already included in our definition of the
light cone quark distribution function.  At high $q^2$, our general
(all twist) result for $F_2$ agrees with the leading twist result
derived using Eq.~\ref{ltf2}~\cite{MV99}.

Clearly, to compute $W^{\mu\nu}$, we first need to compute the fermion
Green's function in the classical background field.  The field
strength carried by these classical gluons is highly singular, being
peaked about the source (corresponding to the parton current at x
values larger than those in the field) localized at $x^-=0$. Away from
the source, the field strengths are zero and the gluon fields are pure
gauges on both sides of $x^-=0$ (see Eq.~\ref{ap1}). The fermion
wavefunction is obtained by solving the Dirac equation in the
background field on either side of the source and matching the
solutions across the discontinuity at $x^-=0$. Once the eigenfunctions
are known, the fermion propagator can be constructed, in the standard
fashion, by writing
 \be
S(x,y) = \int {d^4 q \over {(2\pi)^4}} {1\over {q^2 + M^2
-i\varepsilon}}
\sum_{pol} \psi_{q} (x) {\bar\psi}_{q} (y) \, ,
\ee
after identifying $q^+ = (q_t^2+M^2-\lambda)/2q^-$. We will not
discuss the details of the derivation here but refer the interested
reader to Ref.~\cite{MV99}.

Define
\be
        G(x_t,x^-) = \theta(-x^-) + \theta(x^-) V(x_t)\, ,
\ee
a gauge transformation matrix that transforms the gluon
field at hand to a singular field with the only non--zero component,
$A^{\prime \mu} = \delta^{\mu +}\alpha(x_t)$. Our result then is that the
fermion propagator in the background field has the form~\cite{MV2,MV99}
\be
        S_A(x,y) &  = & G(x)S_0(x-y)G^\dagger(y)
             -i  \int d^4 z G(x)\Bigg\{ \theta(x^-)\theta(-y^-)
                     ( V^\dagger(z_t)-1) - \nonumber \\
                     & & \theta(-x^-)\theta(y^-)(V(z_t)-1)\Bigg\}
G^\dagger(y)
               S_0(x-z) \gamma^- \delta(z^-) S_0 (z-y) \, .
\label{nsprop}
\ee
with the free fermion Green's function 
\be
S_0 (x-y)=\int {d^4q\over {(2\pi)^4}}
e^{iq\cdot (x-y)}{(M-q\!\!/)\over{q^2+M^2-i\varepsilon}} \,\,\, .
\ee
Recall that $V(x_t)$ is the gauge transformation matrix in the
fundamental representation and that the classical solution $A^i =
V(x_t)\partial^i V^\dagger (x_t)/(-ig)$.  This very simple form of the
propagator is useful in the manipulations below.

In fact, since the current-current correlation function is explicitly gauge
invariant, we may use the singular gauge form of the	
propagator~\cite{HW,Balitsky} for computing the current-current 
correlation function
\be
        S_A^{sing}(x,y)&  = & S_0(x-y) -i \int d^4z 
\Bigg\{\theta(x^-)\theta(-y^-)
(V^\dagger(z_t)-1) - \nonumber \\
& &  \theta(-x^-)\theta(y^-)(V(z_t)-1) \Bigg\} S_0(x-z)\gamma^- \delta(z^-)
S_0(z-y)\, .
\label{singprop}
\ee
A diagrammatic representation of the form of the propagator above is
shown in Fig.~1
In the expressions below for $W^{\mu \nu}$ we will drop the superscript
$sing$ and simply use the singular gauge expression for the propagator.
\vskip -0.05in
\begin{figure}[ht]
\begin{center}
\setlength\epsfxsize{6in}
\epsffile{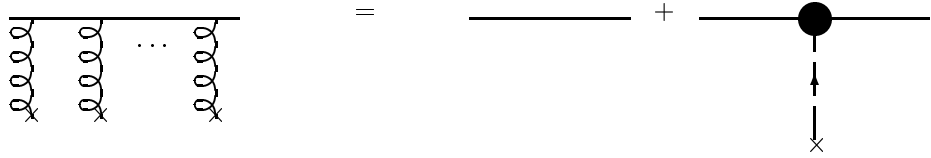}
\caption{Diagrammatic representation of the propagator in Eq.~48.}
\end{center}
\end{figure}

Our result for the fermion propagator in the classical background
field was obtained for a $\delta$--function source in the $x^-$
direction. This assumption was motivated by the observation that small
$x$ modes with wavelengths greater than $1/P^+$ perceive a source which
is a $\delta$-function in $x^-$. The propagator above can also be
derived for the general case where the source has a dependence on
$x^-$. The gauge transforms above are transformed from
$V(x_t)\rightarrow V(x_t,x^-)$, to path ordered exponentials, where
$V(x_t,x^-)$ is given by Eq.~\ref{puresoln}. Our result for the
propagator is obtained as a smooth limit of $\Delta x^- = 1/xP^+ >>
x^- (=1/P^+)$.  Therefore our form for the propagator is the correct
one provided we interpret the $\theta$-functions and
$\delta$-functions in $x^-$ to be so only for distances of interest
greater than $1/P^+$, the scale of the classical source.

We are now in a position to calculate the current--current correlator.
This calculation is accurate
to lowest order in $\alpha_S$ but to all orders in $\alpha_S\mu$.
Before we go ahead with the computation, we will discuss briefly 
the averaging procedure over the labels of color charges at 
rapidities higher 
than those of interest. This is required if we are to compute gauge invariant 
observables. 

If we average the Green's function in Eq.~\ref{nsprop} over all
possible values of the color labels corresponding to the partons at
higher rapidities, we can employ the following definitions for future
reference.  Defining
\be
{1\over N_c}\,<{\rm Tr}\left(V (x_t) V^\dagger (y_t)\right) >_{\rho} =  
\gamma (x_t - y_t) \, ,
\ee
we see that $\gamma (0) = 1$, 
which follows from the unitarity of the matrices $V$.  Now defining
the Fourier transform\footnote{We define the Fourier transform in
this way because it corresponds to only the connected pieces in the
correlator.}
\be
        {\tilde\gamma} (p_t) = \int d^2x_t~ e^{-ip_tx_t} 
\left[\gamma(x_t)-1\right] \, ,
\label{Fourgamm}
\ee
we have the sum rule
\be
        \int {{d^2p_t} \over {(2\pi )^2}}~ {\tilde \gamma}(p_t) = 0 \, .
\label{rule}
\ee
The function ${\tilde \gamma} (p_t)$ will appear frequently in our
future discussions and as we shall see, can be related to the gluon
density at small $x$.

We will now use the fermion Green's function in Eq.~\ref{nsprop} to
derive an explicit result for the hadronic tensor $W^{\munu}$. 
In the following section, we will compute the structure
functions $F_1$ and $F_2$.
As previously, we define
\be
W^{\munu} (q, P, X^-) = {\rm Im}\, \int d^4 z\, e^{iq\cdot z}\,
<T(J^\mu (X^- + {z\over 2})J^\nu (X^- - {z\over 2})> \, ,
\ee
where ``Imaginary'' stands for the discontinuity in $q^-$. Then
\be
& &W^{\munu} (q, P) = {1\over 2\pi}\sigma {P^+ \over m} 
\int dX^- \, W^{\munu} (q, P, X^-)
\equiv {1\over 2\pi}\sigma P^+ {\rm Im}\,\int dX^- \int d^4 z\, e^{iq\cdot z} \nonumber \\
&\times& {\rm Tr}\left(
S_{A_{cl}}\left(X^- + {z\over 2}, X^- -{z\over 2}\right)\,\gamma^\nu
S_{A_{cl}}\left(X^- - {z\over 2}, X^- +{z\over
2}\right)\,\gamma^\mu\right) \, .
\ee
The only terms in the propagator that contribute to the above are the
$\theta(\pm x^-)\theta(\mp y^-)$ pieces. Using our representation for
the propagator in Eq.~\ref{singprop}, after considerable
manipulations, we can write $W^{\munu}$ as
\be
W^{\munu}(q,P) &=& {{\sigma P^+ N_c} \over {2\pi m}} {\rm Im} \int {d^4 p
\over
{(2\pi)^4}}\,{d^2 k_t \over {(2\pi)^2}}{dk^+\over {(2\pi)}}
\,\,{\tilde\gamma}(k_t) \nonumber \\
&\times&{\rm Tr} \left\{ (M-p\!\!/)\gamma^- (M-l\!\!/)\gamma^\mu
(M-{l^\prime}\!\!/)\gamma^- (M-{p^\prime}\!\!/)\gamma^\nu \over
{ (p^2 + M^2 -i\varepsilon)\,(l^2 + M^2 -i\varepsilon)\,
({l^\prime}^2 + M^2 -i\varepsilon)\,({p^\prime}^2 + M^2 -i\varepsilon)}
\right\}\, , \nonumber \\
\label{wmunu}
\ee
where $l=p-k$, $l^\prime = l-q$, $p^\prime = p-q$ and $k^-=0$.
Correspondingly,  we can write $W^{\munu}$ as the imaginary
part of the diagram shown in Fig.~2.

\begin{figure}[ht]
\begin{center}
\setlength\epsfxsize{4in}
\epsffile{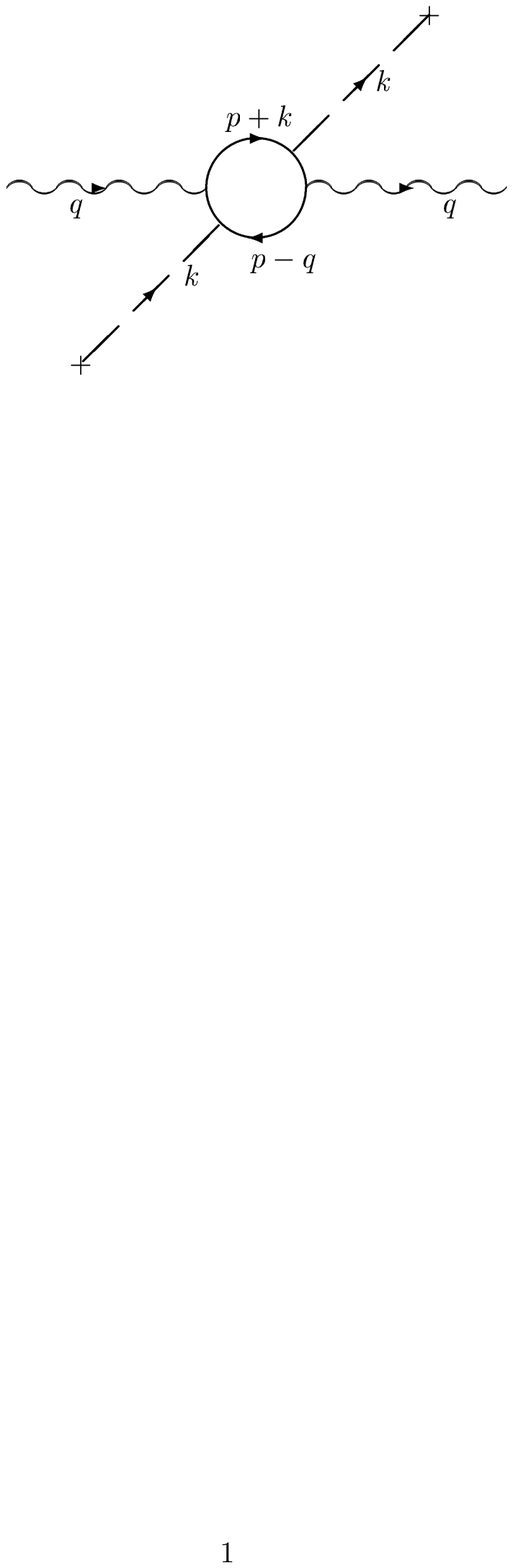}
\caption{Polarization tensor with arbitrary number of insertions from the
classical background field. Wavy lines are photon lines, the solid
circle denotes the fermion loop and the dashed lines are the insertions
from the background field (see Fig.~1). The imaginary part of this diagram
gives $W^{\munu}$.}
\end{center}
\end{figure}

For the DIS case, $q^2 > 0$ (see footnote~1), 
we can cut the
above diagram only in the two ways shown in Fig.~3 (the diagram where both
insertions from the external field are on the same side of the cut
is forbidden by the kinematics).

\begin{figure}[ht]
\setlength\epsfxsize{6in}
\epsffile{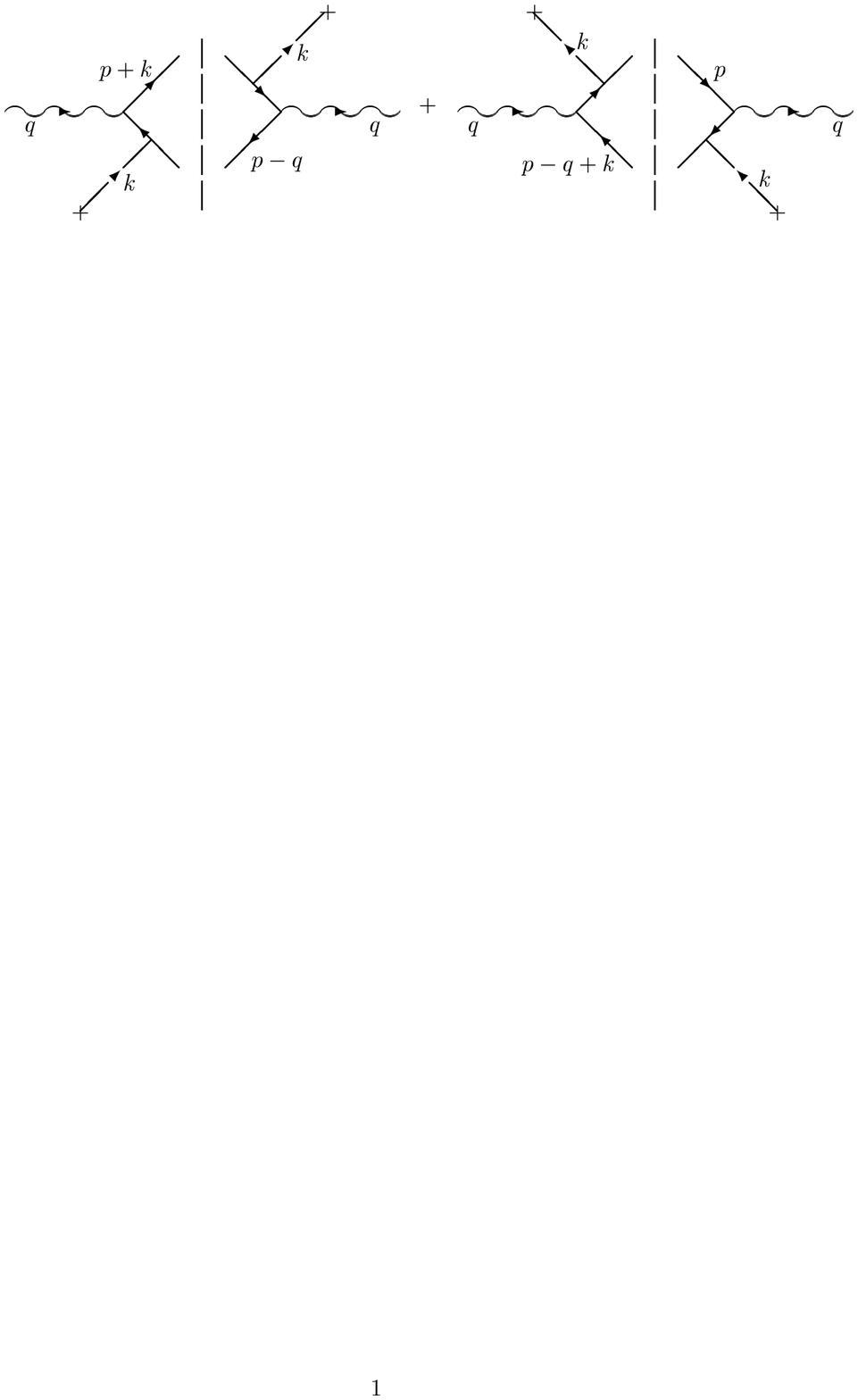}
\caption{Cut diagrams corresponding to the imaginary part of $W^{\munu}$.}
\end{figure}

Also interestingly, the contribution to $W^{\munu}$ can be represented
solely by the diagram in Fig.~4 and not, as is usually the case, from
the sum of this diagram and the standard box diagram. This is because
in our representation of the propagator multiple insertions from the
external field on a quark line can be summarized into a single
insertion. Eq.~\ref{singprop} makes this point clear.

\begin{figure}[ht]
\setlength\epsfxsize{5.5in}
\epsffile{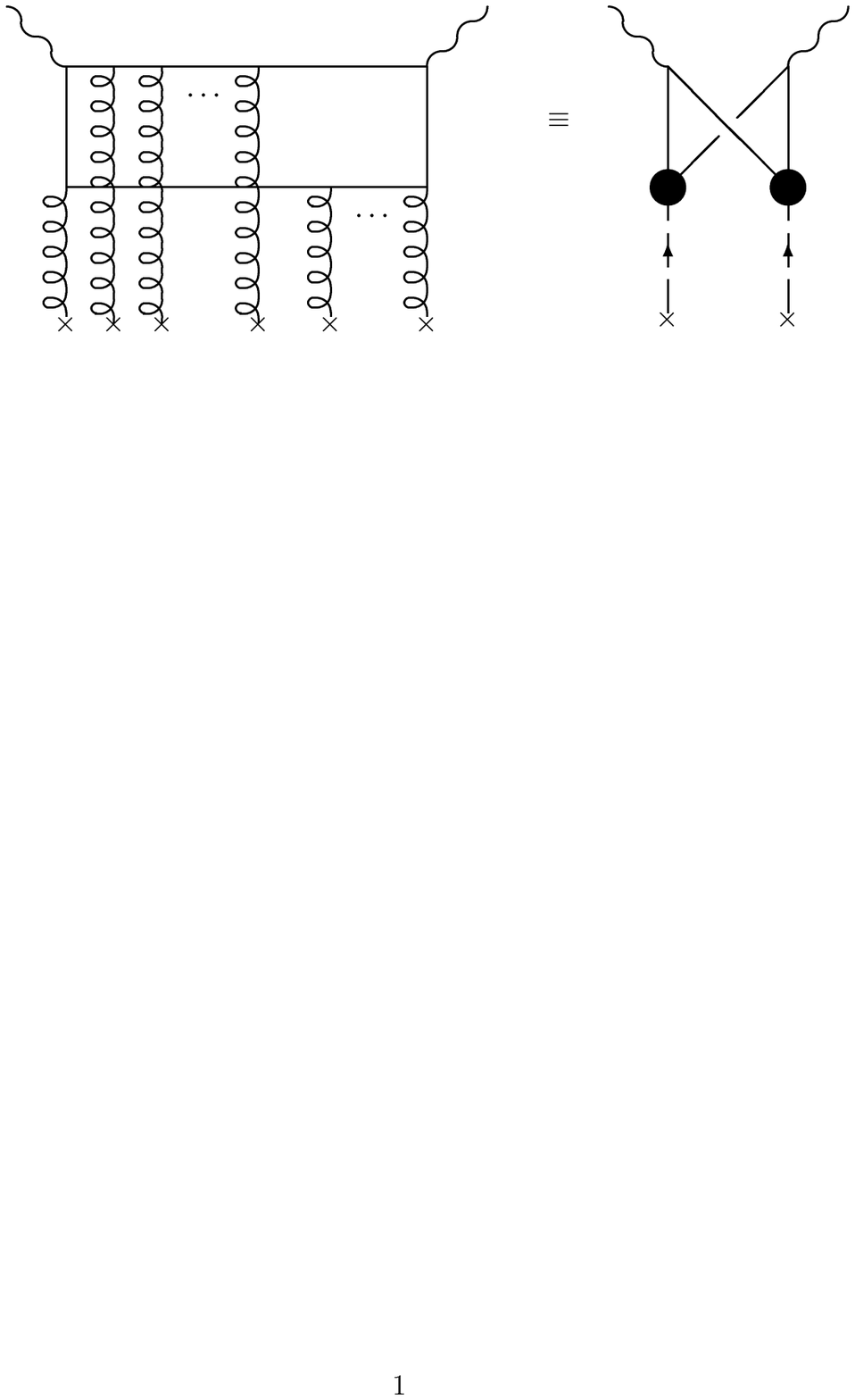}
\caption{In the singular gauge representation for the propagator (see 
Eq.~\ref{singprop} and Fig.~1), the current--current correlator
(imaginary part of LHS) is equivalent to the imaginary part of RHS.}
\end{figure}

Applying the Landau--Cutkosky rule, shifting $p\rightarrow
p+k$,  and changing variables appropriately, 
Eq.~\ref{wmunu} can be written as
\be
& &W^{\munu}(q,P) = {\sigma P^+ N_c\over {2\pi m}}\int {d^4 p \over
{(2\pi)^4}}\,{d^2 k_t \over {(2\pi)^2}}{dk^+\over {(2\pi)}}
\,\,{\tilde\gamma}(k_t) M^{\munu}\theta(p^+ + k^+)\theta(-p^+)\nonumber
\\
&\times& (2\pi)^2
\delta((p+k)^2 + M^2)\,\delta((p-q)^2 + M^2))
{1\over {p^2 + M^2}}\,{1\over {(p+k-q)^2 + M^2}}\, ,\nonumber\\
\ee
where above the trace is represented by~\footnote{Kinematic note: the 
observant reader will notice we have put $q^+=0$ here. Since 
we are working in the infinite momentum frame, the hadron has only one 
large momentum component, $P^+$. The rest are put to zero. For the photon, 
we choose a left moving frame such that $q^0 = |q^z|$ and $q^+=0$. Then, 
$q^2 = q_t^2 >0$, $P\cdot q= -P^+ q^-$ and $x_{Bj} = -q^2/(2P\cdot q)
\equiv q_t^2/(2P^+ q^-)$. Since in the infinite momentum frame 
$ 0<x_{Bj}<1 $, this gives $ q^- > 0$.} 
\be
M^{\munu} = {\rm Tr}
\Bigg\{ (M-p\!\!/-k\!\!/)\gamma^- (M-p\!\!/)\gamma^\mu
(M-p\!\!/+q\!\!/)\gamma^- (M-p\!\!/-k\!\!/+q\!\!/)\gamma^\nu
+ \mu\leftrightarrow \nu \Bigg\} \, .\nonumber\\
\ee

In Ref.~\cite{MV99}, the integral over $p^-$ in Eq.~\ref{wmunu} was 
performed before the $p^+$ 
integral. Here, we instead perform the $p^+$ integral first. Further, defining 
$z= p^-/q^-$, we note that the $\theta$--function and $\delta$--function 
constraints in Eq.~\ref{wmunu} restrict $0<z<1$. This simplifies the result 
in Ref.~\cite{MV99} considerably. Performing the $p^+$ integral, one obtains 
\be
& &W^{\munu}(q,P) = {\sigma P^+ q^- N_c\over {\pi^2 m}}\,{1\over (q^-)^2}
\int_0^1\,dz\,\int\,{d^2p_t\over (2\pi)^2}\,{d^2 k_t \over {(2\pi)^2}}
\,{\tilde\gamma}(k_t) {M^{\munu}\over 16}\nonumber
\\
&\times& 
{1\over {M_p^2+z(1-z)q_t^2}}\,\,\,{1\over {M_{p+k-q}^2+z(1-z)q_t^2}}\, 
,\nonumber\\
\label{finwmunu}
\ee
where $M_p^2 = p_t^2 + M^2$. Similarly $M_{p+k-q}^2=(p_t+k_t-q_t)^2+M^2$.
The above is the final result of this section, and will be used below in 
computing structure functions.

\section{Structure Functions at Small x}
\vskip 0.1in

The hadronic tensor $W^{\munu}$ can be decomposed in terms of the
structure
functions $F_1$ and $F_2$ as~\cite{Pokorski}
\be
m W^{\munu} = -\Bigg(g^{\munu} - {q^\mu q^\nu \over {q^2}}\Bigg) F_1
+ \Big(P^\mu - {q^\mu (P\cdot q)\over {q^2}}\Big)\,\Big(P^\nu -
{q^\nu {P\cdot q}\over {q^2}}\Big) {F_2\over {(P\cdot q)}} \, , 
\label{strctfn}
\ee
where $P^\mu$ is the four--momentum of the hadron or nucleus and
$P^2= m^2 \approx
0$ ($<<q^2$). In the infinite momentum frame, we have $P^+\rightarrow
\infty$ and $P^-, P_t \approx 0$.
The above equation can be inverted to obtain expressions for $F_1$
and $F_2$ in terms of components of $W^{\munu}$. 
Since in our kinematics $q^+=0$ (see footnote 3 for a kinematic note) we have 
\be
F_1 = {F_2\over 2x} + \left({q^2 \over {(q^-)}^2}\right)\, W^{--} \,\,\,;\,\,\,
{1\over 2x} F_2 = -\left( {(q^-)^2 \over q^2}\right)\, W^{++} \, .
\ee

It is useful to verify explicitly that our expression for $W^{\mu \nu}$
derived in an external field can be written in the form of Eqn. \ref{strctfn}.
Recall that $W^{\mu \nu}$ can be written in Lorentz covariant form 
by using the vector $n^\mu = \delta^{\mu +}$.  Using $n \cdot \gamma =
-\gamma^-$ in Eqn. \ref{wmunu}, we see that $W^{\mu \nu}$ is a Lorentz 
covariant function of the only vectors in the problem--$q^\mu$ and $n^\mu$.
Identifying $n^\mu = P^\mu /P^+$ in Eqn. \ref{strctfn}, we see that these
forms are in complete agreement.  Note that all factors of $m$
disappear from $F_1$ and $F_2$ by the explict forms of Eqns. \ref{strctfn}
and \ref{wmunu}.  

We also see that the structure functions can only be functions of $q^2$
and $n\cdot q$ by Lorentz invariance.  We can therefore take $q^+ = 0$
for the purpose of computing $F_1$ and $F_2$.

To compute $W^{++}$ and $W^{--}$, we need to know the
the traces $M^{++}$ and $M^{--}$, respectively in Eq.~\ref{finwmunu}.
We can compute them explicitly and the results can be represented 
compactly as
\be
{1\over 16} M^{++} = {1\over 2}\left( M_p^2 M_{p+k-q}^2 + M_{p+k}^2 
M_{p-q}^2 - q_t^2 k_t^2 \right) \, ,
\ee      
and
\be
M^{--} = 32 (p^-)^2 (p^- - q^-)^2 \, . 
\ee
From the relations above of $F_1$ and $F_2$ to $W^{++}$ and $W^{--}$, 
we obtain from Eq.~\ref{finwmunu} the following general results for 
the structure functions for arbitrary values of $Q^2$, $M^2$ and the
intrinsic scale $\mu$,\footnote{which is implicitly contained in the
function ${\tilde \gamma(k_t)}$ in Eq.~\ref{finwmunu}.}
\be
F_2 = &=& {Q^2\sigma N_c\over {2\pi^3}} 
\int_0^1 dz \int_0^{1\over \Lambda_{QCD}} 
dx_t\, x_t \,\left(1-\gamma(x_t,y)\right) \, \nonumber \\
&\times& \left[K_0^2 (x_t A)\left(4z^2(1-z)^2 Q^2 +M^2\right)+ K_1^2(x_t A)
A^2 \left(z^2+(1-z)^2\right)\right] \, .
\ee
Here $A^2 = Q^2 z(1-z) + M^2$ and $K_{0,1}$ are the modified Bessel
functions. For simplicity, we have ignored the impact parameter
dependence of $\gamma$--and replaced the integral over impact parameter
by the transverse area $\sigma$. For the same reason, we ignore the
sum over the charge squared of the quark flavors. Both of these must
of course be included in numerical computations. The first (second)
term in the bracket above is the probability for a longitudinally
(transversely) polarized photon to split into a ${\bar q} q$ pair.
Ignoring target mass corrections which are negligible at small $x$, 
\be
F_L &=& F_2 - 2x F_1 \nonumber \\
&\equiv&   {Q^2\sigma N_c\over {\pi^3}} 
\int_0^1 dz \int_0^{1\over \Lambda_{QCD}} 
dx_t\, x_t \,\left(1-\gamma(x_t,y)\right)\,z^2(1-z)^2 Q^2\, K_0^2 (x_t A) \, .
\label{generalf1}
\ee

For a Gaussian source (see Ref.~\cite{MV99} and footnote 1), 
\be
\gamma(x_t,y) = 
\exp\left(- {\alpha_S \pi^2\over {2\sigma N_c}}x_t^2 x G(x,{1\over 
x_t^2})\right) \, ,
\label{effvertex}
\ee
where the scale is set by the transverse separation $x_t$ between the 
quark and the anti--quark. 

The equation for $F_2$ with the Gaussian source is the well known
Glauber expression~\cite{NikZak} usually derived in the rest frame of
the nucleus. It is heartening to see that the formalism of
Ref.~\cite{MV99} for structure functions in the infinite momentum
frame reproduces it.  For large $Q^2$, it reduces to the standard
small $x$ DGLAP expression\footnote{In Ref.~\cite{MV99}, it was shown
explicitly that our general expression for $F_2$ formally reduces to
the leading twist expression obtained from Eq.~\ref{distprop}.} while
at small $Q^2$ it goes to zero as $Q^2 log(Q^2)$. One then recovers,
qualitatively, the shape of the famous Caldwell plot for
$dF_2/d\log(Q^2)$ measured at HERA~\cite{Caldwell}. Similar forms 
were used by several authors to understand the recent data~\cite{GLMBW}.

One obtains from the above equation for $F_2$, in a manner analogous to
Eq.~\ref{glue}, the quark saturation scale $Q_s^q$ by replacing
$C_F\longrightarrow C_A$ in Eq.~\ref{gluesat}.  The relative size of
the two saturation scales, glue to quark, is therefore determined
simply by the ratio of the two Casimirs, $C_A/C_F$.

What about quantum corrections to the above quark and gluon
distributions?  At the one loop level, one gets $\log(1/x)$
corrections to the Weizsacker--Williams
distribution~\cite{Muell90,AJMV,Muell991}. However, Mueller has argued
recently that beyond the one loop level, the distribution has the same
form as the as the above classical gluon distribution. What does
change due to small $x$ evolution is the $x$ dependence of the
saturation scale~\cite{Muell991}.  Recently, there have been detailed studies 
by Kovchegov, and by Levin and Tuchin, of parton evolution in the 
non--linear region~\cite{Yuri99,Levin}. Their results appear to confirm the intuitive picture 
of Mueller.

As $q^2\longrightarrow \infty$, we find remarkably that
the integral on the RHS of Eq.~\ref{generalf1} vanishes and 
it reduces to
\be
F_1 = {F_2\over 2x} \, .
\ee
The above is the well known Callan--Gross relation.
The reader 
may note above that the deviation from the Callan--Gross relation vanishes 
as a power law as $q^2\rightarrow \infty$. On the other hand, it is well 
known in QCD~\cite{ZWT,BBDM} that the violations of the 
Callan--Gross relation only disappear logarithmically 
as $q^2\rightarrow \infty$. The apparent contradiction is resolved by 
one realizing that the logarithmic violations  at large $q^2$ in QCD come from 
diagrams where the sea quark emits a gluon (thereby violating Feynman's parton 
model helicity argument). These diagrams are of higher order in our picture 
and are therefore not included. In fact, the deviations from the
Callan--Gross relation of the sort discussed above (at small x) should die 
off faster than logarithmically at very large $q^2$ because for sufficiently 
large $q^2$, the violations of the Callan--Gross relation should come 
from precisely the diagrams not included here. At moderate $q^2$ however, the
contributions we have discussed above should be important.

\section{The non--Abelian Weizs\"acker--Williams approach to high energy 
nuclear collisions}
\vspace*{0.1cm}

In nuclear collisions at very high energies, the hard valence parton modes
in each of the nuclei 
act as highly Lorentz contracted, static sources of color charge for the
wee parton, Weizs\"acker--Williams modes in the nuclei. The sources are
described by the current
\be
J^{\nu,a}(r_t) = \delta^{\nu +}g\rho_{1}^a (r_t)\delta(x^-) + \delta^{\nu -}
g\rho_{2}^a (r_t) \delta(x^+) \, ,
\label{sources}
\ee
where $\rho_1$ ($\rho_2$) correspond to the color charge densities of
the hard modes in nucleus 1 (nucleus 2) respectively.  The classical
field of two nuclei is described by the solution of the Yang--Mills
equations in the presence of the light cone sources:
\be
D_\mu F^{\mu\nu} = J^\nu \, ,
\label{yangmill}
\ee

Gluon distributions are simply related to the Fourier transform $A_i^a (k_t)$ 
of the solution to the above equation by $<A_i^a(k_t) A_i^a(k_t)>_\rho$. The
averaging over the classical charge distributions is defined by
\be
\langle O\rangle_\rho = \int d\rho_{1}d\rho_{2}\, O(\rho_1,\rho_2)
\,\exp\left( -\int d^2 r_t {{\rm Tr}\left[\rho_1^2(r_t)+\rho_2^2(r_t)
\right]
\over {2\mu^2}}\right) \, ,
\ee
and is performed independently 
for each nucleus with equal Gaussian weight $\mu^2$. Of course, this is
only true for identical nuclei.

Before the nuclei collide ($t<0$), a solution of the equations of motion is
\be
A^{\pm}=0 \,\,\,;\,\,\,
A^i= \theta(x^-)\theta(-x^+)\alpha_1^i(r_t)+\theta(x^+)\theta(-x^-)
\alpha_2(r_t) \, ,
\label{befsoln}
\ee
where $\alpha_{q}^i(r_t)$ ($q=1,2$ denote the labels of the nuclei) 
are pure gauge fields of the two nuclei before the collision and have the 
form described in Eq.~\ref{puresoln}.
The above expression suggests that for $t<0$ the solution is simply the
sum of two disconnected pure gauges.

For $t>0$ the solution is no longer pure gauge. Working in the Schwinger 
gauge
$A^\tau\equiv  x^+ A^- + x^- A^+ =0$
the  authors of Ref.~\cite{KLW} found that with the ansatz
\be
A^{\pm}=\pm x^{\pm}\alpha(\tau,r_t)\,\,\,\, ;\,\,\,\,
A^i =\alpha_\perp^i(\tau,r_t) \, ,
\label{ansatz}
\ee
where $\tau=\sqrt{2x^+ x^-}$, Eq.~\ref{yangmill} could be written in
the simpler form
\be
{1\over \tau^3}\partial_\tau \tau^3 \partial_\tau \alpha + [D_i,\left[D^i,
\alpha\right]]
&=&0 \, , \nonumber \\
{1\over \tau}[D_i,\partial_\tau \alpha_\perp^i] + ig\tau[\alpha,\partial_
\tau \alpha] &=&0\, ,\nonumber \\
{1\over \tau}\partial_\tau \tau\partial_\tau \alpha_\perp^i
-ig\tau^2[\alpha,\left[D^i,\alpha\right]]-[D^j,F^{ji}]&=&0 \, . 
\label{yangmill2}
\ee 
The above equations of motion are independent of $\eta$--the 
gauge fields in the forward light cone are therefore only functions of
$\tau$ and $r_t$ and are explicitly boost invariant. This result follows from 
the assumption that the sources of color charge are delta functions on the 
light cone. Of course this is not true in general. However, we are interested 
in the region of central rapidity, about one unit of rapidity around 
$\eta=0$. The boost invariance assumption should be Ok in this region. Also 
note that boost invariance is {\it not} assumed when solving for the fields of 
the nuclei before the collision.

The initial conditions for the fields $\alpha(\tau,r_t)$ and $\alpha_\perp^i$
at $\tau =0$ are obtained by matching the equations of motion
(Eq.~\ref{yangmill}) at the point $x^\pm =0$ and along the boundaries
$x^+=0,x^->0$ and $x^-=0,x^+>0$. 
Remarkably, there exist a set of non--singular initial
conditions for the smooth evolution of the classical 
fields in the forward light
cone. These can be written in terms of the fields of each of the nuclei
before the collision ($t<0$) as follows,
\be
\alpha_\perp^i|_{\tau=0}= \alpha_1^i+\alpha_2^i \,\,\,;\,\,\,
\alpha|_{\tau=0}&=&{ig\over 2} [\alpha_1^i,\alpha_2^i] \, .
\label{initial}
\ee
Gyulassy and McLerran have shown~\cite{gyulassy} that even when the
fields $\alpha_{1,2}^i$ before the collision are smeared out in
rapidity to properly account for singular contact terms in the
equations of motion the above boundary conditions remain unchanged.
Further, the only condition on the derivatives of the fields that
would lead to regular solutions are $\partial_\tau
\alpha|_{\tau=0},\partial_\tau \alpha_\perp^i |_{\tau=0} =0$.

In Ref.~\cite{KLW}, perturbative solutions (for small $\rho$) were
found to order $\rho^2$ by expanding the initial conditions and the
fields in powers $\rho$ (or equivalently, in powers of
$\alpha_S\mu/k_t$) We will not discuss the details of the perturbative
solution but wish to refer the reader to the original papers.

Perturbatively, at late times, the fields in the forward light cone can 
be expanded out in plane waves. 
The energy distribution in a transverse box of size $R$ 
and longitudinal extent $dz$ can be computed by summing over the energy of 
the modes in the box with the occupation number of the modes given by
the mode functions $a_i(k_t)$. We have then
\be
{dE\over {dy d^2k_t}} = {1\over {(2\pi)^2}}\sum_{i,b} |a_i^b(k_t)|^2 \, .
\ee
The multiplicity distribution of classical gluons is 
defined as $dE/dyd^2 k_t/\omega$. After performing the averaging over the
Gaussian sources, the number distribution of classical gluons is
\be
{dN\over {dyd^2 k_t}} = \pi R^2 {2g^6 \mu^4\over {(2\pi)^4}} {N_c (N_c^2-1)
\over k_t^4} L(k_t,\lambda) \, ,
\label{GunBer}
\ee
where $L(k_t,\lambda)$ is an infrared divergent function at the scale
$\lambda$. This result
agrees with the quantum bremsstrahlung formula of Gunion and
Bertsch~\cite{GunionBertsch} and with several later
works~\cite{DirkYuri,gyulassy,SerBerDir,Guo}.  

The function $L(k_t,\lambda)$ arises 
from long range color correlations that are cut-off either by a nuclear 
form factor (as in Refs.~\cite{GunionBertsch,DirkYuri}), by dynamical screening
effects~\cite{GyuWang,EskMullWang} or in the classical Yang--Mills case 
of Ref.~\cite{KLW}, non--linearities that become large at the scale 
$k_t\sim \alpha_S\mu$. In the classical case,
$L(k_t,\lambda) = \log(k_t^2/\lambda^2)$, 
where $\lambda = \alpha_S\mu$. The formalism used in all these
derivations breaks down at small momenta and one cannot distinguish
between the different parametrizations of the nuclear form
factors. However, at sufficiently high energies, the behaviour of
$L(k_t,\lambda)$ in the infrared is given by higher order (in
$\alpha_S\mu/k_t$) non--linear terms in the classical effective
theory. We hope to understand in the near future how non--perturbative
effects in the classical effective theory dynamically change the gluon
distributions at small transverse momenta.
 
While the Yang--Mills equations discussed above can be solved perturbatively
in the limit $\alpha_S \mu \ll k_t$, it is unlikely that a simple analytical 
solution exists for Eq.~\ref{yangmill} in general. The classical solutions
have to be determined numerically for $t>0$. The straightforward procedure 
would be to discretize Eq.~\ref{yangmill} but it will be more convenient 
for our purposes to construct the lattice Hamiltonian and obtain the lattice
equations of motion from Hamilton's equations. 

Let us first consider the continuum Hamiltonian~\cite{AlexRaj1}.
In the appropriate $(\tau,\eta,x_t)$ co--ordinates, 
the metric is diagonal with 
$g^{\tau\tau}=-g^{xx}=-g^{yy}=1$ and $g^{\eta\eta}=-1/\tau^2$. 
After a little algebra, the Hamiltonian can be written as~\cite{Sasha}
\be
H =\tau\int d\eta {\rm d}^2r_t\left\{{1\over 2} p^{\eta}p^{\eta}
+{1\over {2\tau^2}}p^r p^r + {1\over{2\tau^2}} F_{\eta r}F_{\eta r}
+{1\over 4}F_{xy}F_{xy} + j^\eta A_\eta + j^r A_r\right\} \, .
\label{hamilton}
\ee
Here we have adopted the gauge $A^\tau =0$. 
Also, $p^\eta={1\over \tau}\partial_\tau A_\eta$ and $p^r=\tau \partial_\tau 
A_r$ are the conjugate momenta.

Consider the field strength $F_{\eta r}$ in the above Hamiltonian. If we 
assume approximate boost invariance, or
\be 
A_r (\tau,\eta,\vec{r_t})\approx A_r(\tau,\vec{r_t}); \ \  
A_{\eta}(\tau,\eta,\vec{r_t})\approx \Phi(\tau,\vec{r_t}),
\ee
we obtain  
\be
F_{\eta r}^a = -D_r \Phi^a \, ,
\label{fdstrgth}
\ee
where $D_r =\partial_r -ig A_r$ is the covariant derivative. Further, if we 
express $j^{\eta,r}$ in terms of the $j^{\pm}$ defined in 
Eq.~\ref{sources} we obtain the enormously simplifying result 
that $j^{\eta,r}=0$ for $\tau>0$. Due to boost invariance, our effective 
Hamiltonian acts in 2+1--dimensions. It is possible to relax this assumption, 
but then the numerical simulations are more complicated.

We now consider the equivalent lattice action and Hamiltonian.
The appropriate action is derived
starting from the Minkowski Wilson action in the discretized 4-space and
taking the naive continuum limit in the longitudinal and time 
directions. Replacing $a^2\sum_{zt}$ with
$\int{\rm d}z{\rm d}t$ in the Minkowski Wilson action, we then have for 
the 2+1--dimensional action
\begin{equation}
S=\int{\rm d}z{\rm d}t\sum_\perp\left[{1\over {2N_c}}{\rm Tr}F_{zt}^2
+{1\over N_c}\Re\,{\rm Tr}(M_{t\perp}-M_{z\perp})
-\left(1-{1\over N_c}\Re\,{\rm Tr}U_{\perp}\right)\right],
\label{twodact}
\end{equation}
where
\begin{equation}
M_{t,jn}\equiv {1\over 2}(A_{t,j}^2+A_{t,j+n}^2)-U_{j,n}\left[{1\over 2}
\partial_t^2U^\dagger_{j,n}+
i(A_{t,j+n}\partial_tU^\dagger_{j,n}-\partial_tU^\dagger_{j,n}A_{t,j})
+A_{t,j+n}U^\dagger_{j,n}A_{t,j}\right]\, ,
\label{mterm}
\end{equation}
and similarly for $M_{z,jn}$. 

The equation of motion for a field is obtained by varying $S$ with respect to
that field. For the longitudinal fields $A_{t,z}$ the variation has the usual
meaning of a partial derivative. For transverse link matrices $U_\perp$ the 
variation amounts to a covariant derivative.
Just as in the continuum case, the lattice initial conditions can be 
determined from the lattice action in Eq.~\ref{twodact}. One 
obtains the lattice equations of motion in the four light cone regions and
then 
determines non--singular initial conditions by matching at $\tau=0$ the
coefficients of the most singular terms in the equations of motion.

On the lattice, the initial conditions are the constraints on the
longitudinal gauge potential $A^\pm$ and the transverse link matrices
$U_\perp$ at $\tau=0$.  
The longitudinal gauge potentials can be written as in the continuum case (see
Eq.~\ref{ansatz}) as
\be
A^\pm=\pm x^\pm\theta(x^+)\theta(x^-)\alpha(\tau, x_t)\, .
\label{apm}
\ee
The transverse link matrices are, for each nucleus, pure gauges before the
collision. This fact is reflected by writing
\be 
U_\perp=\theta(-x^+)\theta(-x^-)I+\theta(x^+)\theta(x^-)U(\tau)
+\theta(-x^+)\theta(x^-)U_1+\theta(x^+)\theta(-x^-)U_2 \, ,
\label{uperp}
\ee
where $U_{1,2}$ are pure gauge. The pure gauges are defined on the lattice
as follows. To each lattice site $j$ we assign two 
SU($N_c$) matrices $V_{1,j}$ 
and $V_{2,j}$. Each of these two defines a pure gauge lattice gauge
configuration with the link variables 
$U_{j,\nhat}^q = V_{q,j}V_{q,j+n}^\dagger$ 
where $q=1,2$ labels the two nuclei. As in the continuum, the gauge 
transformation matrices $V_{q,j}$ are determined by the color charge 
distribution $\rho_{q,j}$ of the nuclei, normally distributed with the 
standard deviation $\mu^2$: 
\be
P[\rho_q]\propto\exp\left(-{1\over{2\mu^2}}\sum_j\rho_{q,j}^2\right).
\label{distriblat}
\ee 
Parametrizing $V_{q,j}$ as $\exp(i\Lambda^q_j)$ with Hermitean traceless
$\Lambda^q_j$, we then obtain $\Lambda^q_j$ by solving the lattice
Poisson equation
\be
\Delta_L\Lambda^q_j\equiv\sum_n\left(\Lambda^q_{j+n}+\Lambda^q_{j-n}
-2\Lambda^q_j\right)=\rho_{q,j}.
\label{latpoi}
\ee 
It is easy to verify that the correct 
continuum solution (Eqs.~\ref{befsoln} and \ref{ansatz}) for the transverse 
fields $A_\perp$ is recovered by taking the formal continuum limit of 
Eq.~\ref{uperp}.

The equation of motion for $U_\perp$, contains, upon substitution of
$U_\perp$ from (\ref{uperp}) and $A^\pm$ from (\ref{apm}), singular 
terms containing the product $\delta(x^-)\delta(x^+)$. 
These originate in the double-derivative contributions 
$\Re{\rm Tr}U_\perp^\dagger\partial_+\partial_-U_\perp$ in the action, when
both derivative operators act on the step functions. Since the coefficient
in front of $\delta(x^+)\delta(x^-)$ must vanish in order to satisfy the
equation of motion, a matching relation between $U_\perp$ 
and $U_{1,2}$ is obtained.
\be
{\rm Tr}\,\sigma_\gamma\left[(U_1+U_2)(I+U_\perp^\dagger)
-{\rm h.c.}\right]=0\, .
\label{ucond}
\ee
Our result is that $(U_1+U_2)(I+U_\perp^\dagger)$ should have no
anti-Hermitean traceless part. Note that this condition has the
correct formal continuum limit: writing $U_{1,2}$ as
$\exp(ia_\perp\alpha_{1,2})$ and $U_\perp$ as
$\exp(ia_\perp\alpha_\perp)$, we have, for small $a_\perp$, the result
$\alpha_\perp=\alpha_1+\alpha_2$, as required.  The above condition in
Eq.~(\ref{ucond}) can easily be resolved in the SU(2) case but we have
not yet found a simple closed form expression for $N_c>2$.  For SU(2),
one obtains for the initial condition 
\be
U_\perp=(U_1+U_2)(U_1^\dagger+U_2^\dagger)^{-1} \, .
\label{ucondp}
\ee

For the $A_-$ field, the singularities arise from the Abelian part of
the $F_{+-}^2$ term in the action whose variation with respect to
$A^{+,\gamma}$ gives
\be
{1\over N_c}{\rm Tr}\sigma_\gamma\partial_+(\partial_-A_+-\partial_+A_-) \, .
\ee
Its most singular part is 
$\alpha_\gamma\theta(x^-)\delta(x^+)$.
Varying the $\pm,\perp$ terms (Eq.~\ref{mterm}) in the action
(Eq.~\ref{twodact}) with respect to $A_j^{+,\gamma}$ and selecting the
contributions containing derivatives, one obtains eventually the result
\be
\alpha_\gamma={i\over{4N_c}}\sum_n{\rm Tr}\sigma_\gamma
\left([(U_1-U_2)(U^\dagger-I)-{\rm h.c.}]_{j,n}
-[(U^\dagger-I)(U_1-U_2)-{\rm h.c.}]_{j-n,n}\right)\, .
\label{agamma}
\ee
It is easily seen that the above equation has the correct formal continuum
limit.  
Writing again $U_{1,2}$ as $\exp(ia_\perp\alpha_{1,2})$ and $U$ as
$\exp(ia_\perp\alpha_\perp)$, one finds in the limit of smooth fields,
$\alpha=i\sum_n[\alpha_1,\alpha_2]_n$, as required.

The lattice action is essential to obtain the initial conditions for the 
evolution of fields in the forward light cone. For the evolution, we 
need the lattice Hamiltonian. It is obtained 
by performing a Legendre transform of Eq.~\ref{twodact}
following the standard Kogut-Susskind procedure~\cite{KS}.
The analog of the Kogut--Susskind Hamiltonian here is
\be
H_L&=& {1\over{2\tau}}\sum_{l\equiv (j,\nhat)} 
E_l^{a} E_l^{a} + \tau\sum_{\Box} \left(1-
{1\over 2}{\rm Tr} U_{\Box}\right)  \, ,\nonumber \\
&+& {1\over{4\tau}}\sum_{j,\nhat}{\rm Tr}\,
\left(\Phi_j-U_{j,\nhat}\Phi_{j+\nhat}
U_{j,\nhat}^\dagger\right)^2 +{\tau\over 4}\sum_j {\rm Tr}\,p_j^2,
\label{hl}\ee
where $E_l$ are generators of right covariant derivatives on the group
and $U_{j,\nhat}$ is a component of the usual SU(2) matrices corresponding
to a link from the site $j$ in the direction $\nhat$. The first two terms
correspond to the contributions to the Hamiltonian from the chromoelectric and
chromomagnetic field strengths respectively. In the last equation
$\Phi\equiv \Phi^a\sigma^a$ is the adjoint scalar field with its conjugate
momentum $p\equiv p^a\sigma^a$.

Lattice equations of motion follow directly from $H_L$ of Eq.~\ref{hl}.  For
any dynamical variable $v$ with no explicit time dependence ${\dot
v}=\{H_L,v\}$, where ${\dot v}$ is the derivative with respect to $\tau$, and
$\{\}$ denote Poisson brackets. We take $E_l$, $U_l$, $p_j$, and $\Phi_j$ as
independent dynamical variables, whose only nonvanishing Poisson brackets are
$$\{p_i^a,\Phi_j^b\}=\delta_{ij}\delta_{ab}; \ \ 
\{E_l^a,U_m\}=-i\delta_{lm}U_l\sigma^a; \ \
\{E_l^a,E_m^b\}=2\delta_{lm}\epsilon_{abc}E_l^c$$
(no summing of repeated indices). The equations of motion are consistent with
a set of local constraints (Gauss' laws). 

The results of this section can be summarized as follows. The four
independent dynamical variables are $E_l$, $U_\perp$, $p_j$ and $\Phi_j$.
Their evolution in $\tau$ after the nuclear collision is determined by
Hamilton's equations above and their values at the initial 
time $\tau =0$ are specified by the following initial conditions:
\be
U_\perp |_{\tau=0} &=& (U_1+U_2)(U_1^\dagger + U_2^\dagger)^{-1} \,\, ; 
\,\, E_l |_{\tau=0} = 0 \, . \nonumber \\
p_j |_{\tau=0} &=& 2\alpha \,\, ; \,\, \Phi_j = 0 \, ,
\label{dinitial}
\ee
where $U_\perp$ and $\alpha$ are given by Eq.~\ref{ucondp} and
Eq.~\ref{agamma} respectively. 

\section{Results for gluon production in high energy nuclear collisions}
\vskip 0.1in

In this section we will discuss recent results for the energy density
$\varepsilon$ as a function of the proper time $\tau$~\cite{AlexRaj3}.
Work on computing number distributions is in progress and will be
reported at a later date~\cite{AlexRaj4}. In an earlier work, we
confirmed that, in weak coupling, the results from our numerical
simulations agreed with lattice perturbation theory~\cite{AlexRaj2}.

The computation of energy densities on the lattice is
straightforward. Our main result is contained in
Eq.~\ref{energydensity}. To obtain this result, we compute the
Hamiltonian density on the lattice for each $\rho^\pm$, and then take
the Gaussian average (with the weight $\mu^2$) over between $40$
$\rho$ trajectories for the larger lattices and $160$ $\rho$
trajectories for the smallest ones.

In our numerical simulations, all the relevant physical information is
contained in $g^2\mu$ and $L$, and in their dimensionless product
$g^2\mu L$~\cite{RajGavai}.  The strong coupling constant $g$ depends
on the hard scale of interest; from Eq.~\ref{colordensity}, we see
that $\mu$ depends on the nuclear size, the center of mass energy, and
the hard scale of interest; $L^2$ is the transverse area of the
nucleus. Assuming $g=2$ (or $\alpha_S=1/\pi$), $\mu
=0.5$ GeV ($1.0$ GeV) for RHIC (LHC), and $L=11.6$ fm for
$Au$--nuclei, we find $g^2\mu L\approx 120$ for RHIC and $\approx 240$
for LHC. (The latter number would be smaller for a smaller value of
$g$ at the typical LHC momentum scale.)  As will be discussed later,
these values of $g^2\mu L$ correspond to a region in which one expects
large non--perturbative contributions from a sum to all orders in $Q_s
\sim 6\,\alpha_S\mu$, even if $\alpha_S\ll 1$. (Recall the definition
of the saturation scale in Eq.~\ref{gluesat}.)  Deviations from
lattice perturbation theory, as a function of increasing $g^2\mu L$,
were observed in our earlier work~\cite{AlexRaj2}.

We  shall now discuss some of the results from our numerical simulations. 
In Fig.~5, we plot $\varepsilon\tau/(g^2\mu)^3$,
as a function of $g^2\mu\tau$, in dimensionless units, for the
smallest, largest, and an intermediate value in the range of $g^2\mu
L$'s studied. 
\begin{figure}[h]
\vspace{-5cm}
\centerline{\hspace{-35cm}{\hbox{\psfig{figure=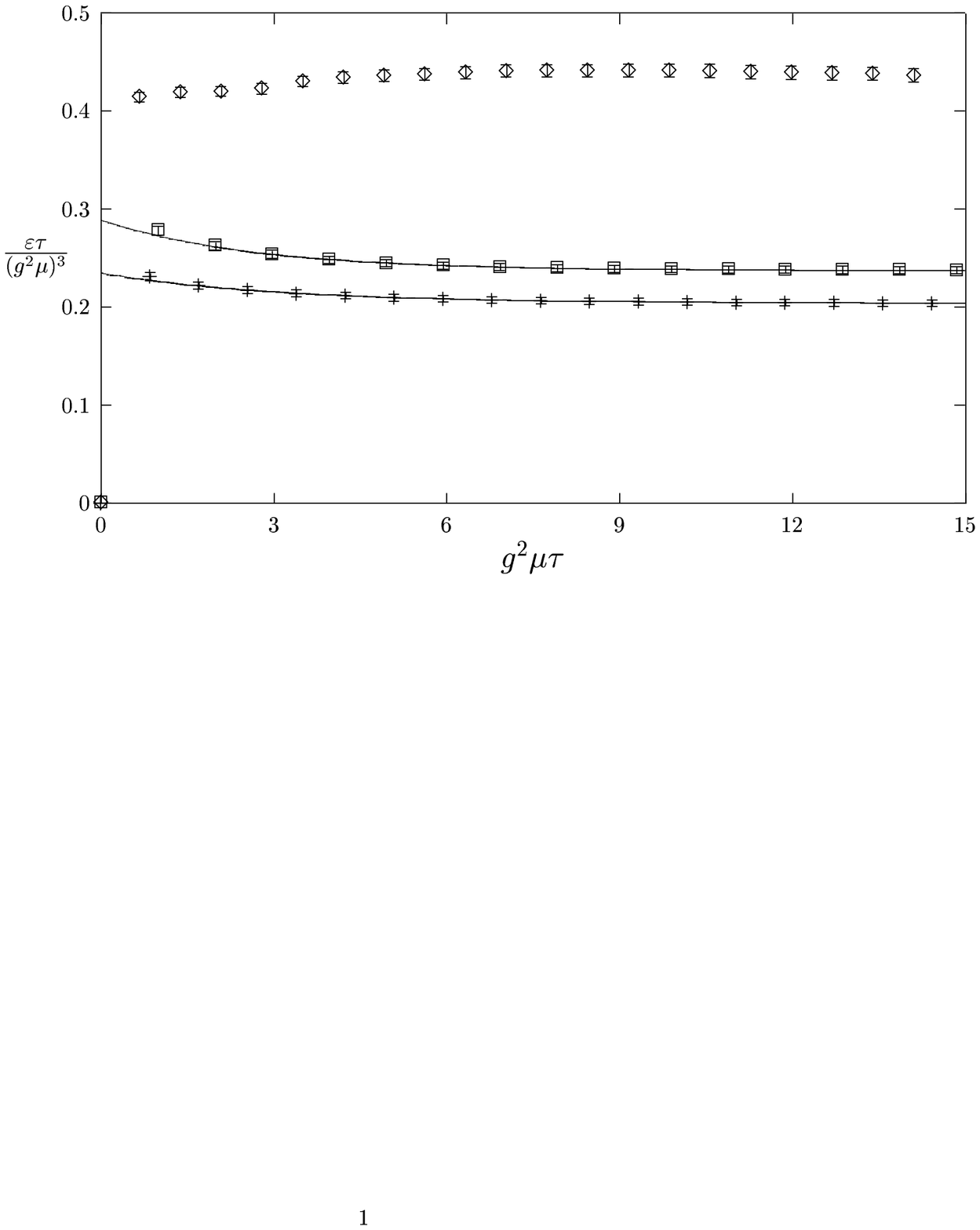,
width=12cm}}}}
\vskip -14cm
\caption{$\varepsilon\tau/(g^2\mu)^3$ as a function of $g^2\mu\tau$
for $g^2\mu L = 5.66$ (diamonds), $35.36$ (pluses)
and $296.98$ (squares). Both axes are in dimensionless units.
Note that $\varepsilon\tau =0$ at $\tau=0$ for all $g^2\mu L$. The lines are
exponential fits $\alpha + \beta\,e^{-\gamma\tau}$ including all
points beyond the peak.}
\label{eXtR}
\end{figure}
The quantity $\varepsilon\tau$ has the physical
interpretation of the energy density of produced gluons $dE/L^2/d\eta$
only at late times--when $\tau\sim t$. Though $\varepsilon \tau$ goes
to a constant in all three cases, the approach to the asymptotic value
is different. For the smallest $g^2\mu L$, $\varepsilon \tau$ increases 
continuously before saturating at late times. 
For larger values of $g^2\mu L$, $\varepsilon\tau$
increases rapidly, develops a transient peak at $\tau\sim 1/g^2\mu$,
and decays exponentially there onwards, satisfying the relation
$\alpha + \beta\,e^{-\gamma\tau}$, to a constant value $\alpha$ (equal
to the lattice $dE/L^2/d\eta$!). The lines shown in the figure are
from an exponential fit including all the points past the peak.  This
behavior is satisfied for all $g^2\mu L \ge 8.84$, independently of $N$.

Given the excellent exponential fit, one can interpret the decay time
$\tau_D=1/\gamma/g^2\mu$ 
as the appropriate scale controlling the formation of gluons
with a physically well defined energy. In other words, $\tau_D$ is the
``formation time''in the sense used by Bjorken~\cite{Bj}.  
In Table~1, we tabulate $\gamma$ versus $g^2\mu
L$ for the largest $N\times N$ lattices for all but the 
smallest $g^2\mu L$. For large
$g^2\mu L$, the formation time decreases with increasing $g^2\mu L$,
as we expect it should. The reason the smallest value of $g^2\mu L$ does not 
have a transient peak is likely because in this case the $k_t$ modes 
do not sufficiently sample the region $k_t\leq Q_s$ where non--linearities 
are important. The few modes there are, lie in the perturbative region where 
the fields can be linearized at $\tau=0$.
\begin{figure}[h]
\vspace{-5cm}
\centerline{\hspace{-35cm}{\hbox{\psfig{figure=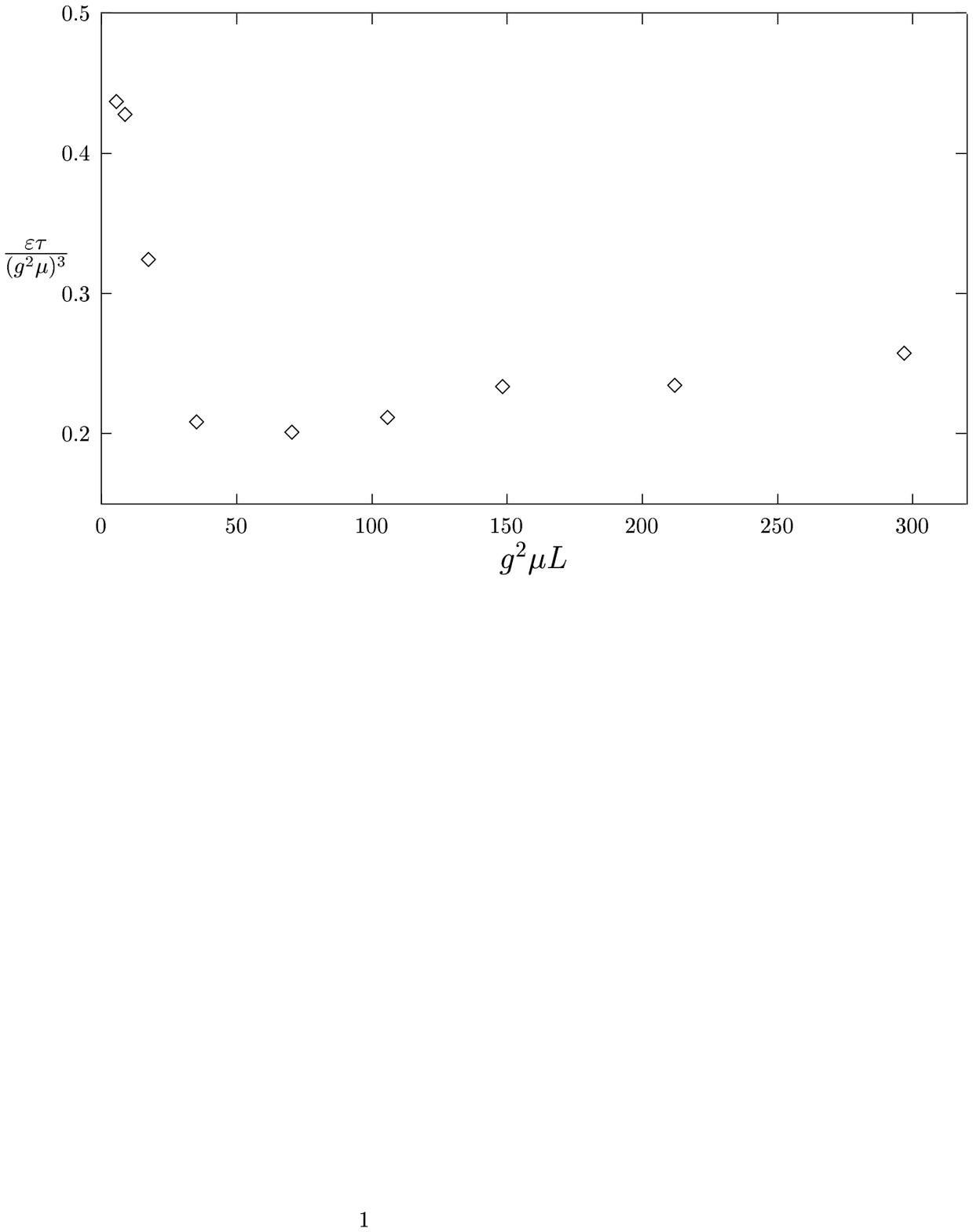,
width=12cm}}}}
\vskip -14cm
\caption{$\varepsilon\tau/(g^2\mu)^3$ extrapolated to the continuum limit: 
$f$ as a function of $g^2\mu L$. The error bars are smaller than the 
plotting symbols.}
\label{eXtvsmuL}
\end{figure}

The physical energy per unit area per unit rapidity of 
produced gluons can be defined in terms of a function $f(g^2\mu L)$ as
\be
{1\over L^2}\, {dE\over d\eta} = {1\over g^2}\,f(g^2\mu L)\,(g^2\mu)^3 \, .
\label{energydensity}
\ee
As discussed in Ref.~\cite{AlexRaj3}, the function $f$ is obtained for 
each fixed value of $g^2\mu L$, by taking the continuum limit, i.e., 
extrapolating $g^2\mu a\longrightarrow 0$.
In Fig.~6, we plot the striking
behavior of $f$ with $g^2\mu L$. For very small $g^2\mu L$'s, it
changes very slightly but then changes rapidly by a factor of two from
$0.427$ to $0.208$ when $g^2\mu L$ is changed from $8.84$ to
$35.36$. From $35.36$ to $296.98$, nearly an order of magnitude in
$g^2\mu L$, it changes by $\sim 25$\%. The precise values of $f$ and
the errors are tabulated in Table~1.
\begin{table}[h]
\centerline{\begin{tabular}{|lrrrrr|} \hline
$g^2\mu L$ & 5.66 & 8.84 & 17.68 & 35.36 & 70.7 \\
$f$ & $.436\pm .007$ & $.427\pm .004$ & $.323\pm .004$ & $.208\pm .004$ 
& $.200\pm .005$ \\
$\gamma$ &  & $.101\pm .024$ & $.232\pm .046$ & $.165\pm .013$ & $.275\pm 
.011$ \\
\hline
$g^2\mu L$ & 106.06 & 148.49 & 212.13 & 296.98 & \\
$f$ & $.211\pm .001$ & $.232\pm .001$ & $.234\pm .002$ & $.257\pm .005$ & \\ 
$\gamma$ & $.322\pm .012$ & $.362\pm .023$ & $.375\pm .038$ & $.378\pm .053$ 
& \\
\hline
\end{tabular}}
\caption{The function $f=dE/L^2/d\eta$ and the relaxation rate 
$\gamma=1/\tau_D/g^2\mu$ 
tabulated as a function of $g^2\mu L$. $\gamma$ has no entry for the 
smallest $g^2\mu L$ since there
$\varepsilon\tau/(g^2\mu)^3$ 
vs $g^2\mu\tau$ differs qualitatively from the other $g^2\mu L$ 
values.}
\end{table}

What is responsible for the dramatic change in the behavior of $f$ as
a function of $g^2\mu L$?  In $A^\tau=0$ gauge, the dynamical
evolution of the gauge fields depends entirely on the initial
conditions, namely, the parton distributions in the wavefunctions of
the incoming nuclei~\cite{KovMueller}.  In the nuclear wavefunction,
at small $x$, non--perturbative, albeit weak coupling, effects become
important for transverse momenta $Q_s\sim 6\,\alpha_s\mu$. 
Now on the lattice, $p_t$ is defined to be $2\pi
n/L$, where $n$ labels the momentum mode.  The condition that momenta
in the wavefunctions of the incoming nuclei have saturated, $p_t\sim
6\,\alpha_S\mu$, translates roughly into the requirement that $g^2\mu L
\geq 13$ for $n=1$.  Thus for $g^2\mu L= 13$, one is only beginning to
sample those modes.  Indeed, this is the region in $g^2\mu L$ in which
one sees the rapid change in $f$.  The rapid decrease in $f$ is likely
because the first non--perturbative corrections are large, and have a
negative sign relative to the leading term. Understanding the later
slow rise and apparent saturation with $g^2\mu L$ requires a better
understanding of the number and energy distributions with $p_t$. This
work is in progress and will be reported on
separately~\cite{AlexRaj4}.

Our results are consistent with an estimate by
A. H. Mueller~\cite{Muell2} for the number of produced gluons per unit
area per unit rapidity. He obtains $dN/L^2/d\eta =
c\,(N_c^2-1)\,Q_s^2/4\pi^2 \,\alpha_S\,N_c$, and argues that the
number $c$ is a non--perturbative constant of order unity. If most of
the gluons have $p_t\sim Q_s$, then $dE/L^2/d\eta =
c^\prime\,(N_c^2-1)\,Q_s^3/4\pi^2\,\alpha_S\,N_c$ which is of the
same form as our Eq.~\ref{energydensity}.  In the $g^2\mu L$ region of
interest, our function $f\approx 0.23$--$0.26$. 
Using the appropriate relation between $Q_s$
and $g^2\mu$, we obtain $c^\prime = 4.3$--$4.9$. 
Since one expects a distribution in momenta about $Q_s$, it is very likely that
$c^\prime$ is at least a factor of $2$ greater than $c$--thereby
yielding a number of order unity for $c$ as estimated by Mueller. This
coefficient can be determined more precisely when we compute the
non--perturbative number and energy distributions.

We will now estimate the initial energy per unit rapidity of produced
gluons at RHIC and LHC energies. We do so by extrapolating from our
SU(2) results to SU(3) assuming the $N_c$ dependence to be
$(N_c^2-1)/N_c$ as in Mueller's formula. At late times, the energy
density is $\varepsilon = (g^2\mu)^4\,f(g^2\mu L)\,\gamma(g^2\mu
L)/g^2$, where the formation time is $\tau_D=1/\gamma(g^2\mu
L)/g^2\mu$ as discussed earlier. We find that $\varepsilon^{RHIC}\approx
66.49$ GeV/fm$^3$ and $\varepsilon^{LHC}\approx 1315.56$
GeV/fm$^3$. Multiplying these numbers by the initial volumes at the
formation time $\tau_D$, we obtain the classical Yang--Mills estimate
for the initial energies per unit rapidity $E_T$ to be $E_T^{RHIC}\approx
2703$ GeV and $E_T^{LHC}\approx 24572$ GeV respectively.
  
We have compared these numbers to results presented recently~\cite{Keijo} 
for the mini--jet energy (computed for 
$p_t > p_{sat}$, where $p_{sat}$ is a saturation scale akin to $Q_s$). He
obtains $E_T^{RHIC} = 2500$ GeV and $E_T^{LHC}=12000$. The remarkable
closeness between our results for RHIC is very likely a
coincidence. The Finnish groups results 
include $K$ factor estimates 
range from $1.5$--$2.5$. If we pick a recent value of 
$K\approx 2$~\cite{Andrei}, we obtain as 
our final estimate, $E_T^{RHIC}\approx 5406$ GeV and $E_T^{LHC}\approx 
49144$ GeV.

To summarize, we discussed in this section 
a non--perturbative, numerical computation,
for a SU(2) gauge theory, of the initial energy, per unit rapidity, of
gluons produced in very high energy nuclear collisions.  Extrapolating
our results to SU(3), we estimated the initial energy per unit
rapidity at RHIC and LHC. We plan to improve our estimates by
performing our numerical analysis for SU(3). Moreover, computations in
progress to determine the energy and number distributions should
enable us to match our results at large transverse momenta to
mini--jet calculations~\cite{AlexRaj4}.

\section{Summary}
\vskip 0.1in

In these lectures, we have discussed a classical effective field
theory approach to scattering at very high energies. At these
energies, a saturation scale $Q_s(x)$ controls the dynamics of high
energy scattering.  How this scale changes as we go to small $x$ is
described by a non--linear renormalization group
equation~\cite{JKW,Yuri99}. The solutions of the RG equations and the
inclusion of effects such as the running of the coupling in the regime
of strong non--linear fields need to be better understood.  In
particular, one would like to investigate possibly striking
experimental signatures of this regime.  Since
$Q_s(x)>>\Lambda_{QCD}$, weak coupling methods may be applicable.  In
these lectures we have shown how one may begin to apply these weak
coupling methods to study DIS and high energy scattering.
 
\section*{Acknowledgements}
\vskip 0.1in
These lectures summarize work done in collaboration with Alex Krasnitz and 
Larry McLerran. I would like to thank Michal Praszalowicz and the other 
organizers of the Zakopane School for their very kind hospitality. This 
work was supported by the Director, Office of Energy Research, Division of 
Nuclear Physics of the Office of High Energy and Nuclear Physics of the 
US Department of Energy under contract number 
DOE-FG02-93-ER-40764.

\end{document}